\documentclass[aps,prb,amsmath,amssymb,superscriptaddress,twocolumn,longbibliography]{revtex4-2}
\usepackage{bbm}
\usepackage{graphicx}
\usepackage{dcolumn}
\usepackage{bm}
\usepackage{subfigure}
\usepackage{amsmath}
\usepackage{stmaryrd}
\usepackage{times}
\usepackage{graphicx}
\usepackage{dcolumn}
\usepackage{color}
\usepackage{makecell}
\usepackage{tabularx,multirow}
\usepackage{appendix}
\usepackage{amssymb}
\usepackage{pifont}
\usepackage[colorlinks,linkcolor=blue,citecolor=blue,urlcolor=black]{hyperref}
\begin{document}
\title{Spin relaxation in $X$-wave magnets with $X=p, d, f, g, i$}
\author{Lijie Liu}
\affiliation{School of Physics, Harbin Institute of Technology, Harbin 150001, China}
\author{Mingbo Dou}
\affiliation{School of Physics, Harbin Institute of Technology, Harbin 150001, China}
\author{Xu Chen}
\affiliation{School of Physics, Harbin Institute of Technology, Harbin 150001, China}
\author{Xianjie Wang}
\affiliation{School of Physics, Harbin Institute of Technology, Harbin 150001, China}
\affiliation{Heilongjiang Provincial Key Laboratory of Advanced Quantum Functional Materials and Sensor Devices, Harbin 150001, China}
\author{M. Ye. Zhuravlev}
\affiliation{Faculty of Liberal Arts and Sciences, St. Petersburg State University, St. Petersburg 190000, Russia}
\author{A. V. Nikolaev}
\affiliation{Skobeltsyn Institute of Nuclear Physics, Moscow State University, Moscow 101000, Russia}
\author{L. L. Tao}
\email{Contact author: lltao@hit.edu.cn}
\affiliation{School of Physics, Harbin Institute of Technology, Harbin 150001, China}
\affiliation{Heilongjiang Provincial Key Laboratory of Advanced Quantum Functional Materials and Sensor Devices, Harbin 150001, China}
\date{\today}
\begin{abstract}
Spin relaxation results in the spin decoherence and a finite spin lifetime, which are detrimental to spintronic devices. To achieve a long spin lifetime desirable for spintronic devices, elucidating the spin relaxation mechanism and factors influencing the spin lifetime is of vital importance. Here, we investigate the spin relaxation in $X$-wave magnets ($X=p, d, f, g, i$) with Rashba spin–orbit coupling within the framework of D'yakonov-Perel' mechanism. We calculate the general matrix of the spin relaxation time for an arbitrary N\'eel vector direction of the $X$-wave magnet. As an illustration, we study the spin relaxation for the N\'eel vector along the $[001]$ direction. It is found that the reciprocal spin-relaxation-time matrices are anisotropic and diagonal for the $d$-, $f$-, $g$- and $i$-wave magnets. For the $p$-wave magnet, we derive the analytical expressions for the temporal evolution of spins. Moreover, the spin relaxation rate is proportional to the momentum relaxation time, Rashba and altermagnetic spin-split strengths for all $X$-wave magnets. Our results shine more light on the fundamental understanding of the spin relaxation mechanism in $X$-wave magnets. 
\end{abstract}
\maketitle
\section{Introduction}
Spins of charge carriers undergo the relaxation processes due to the combined diffuse scattering and spin–orbit coupling (SOC)\cite{pr61,spinbook}. As such, the injected spin lost its polarization or coherence after a certain time known as the spin lifetime, which is a prerequisite for spintronic devices\cite{rmd323,nrm2584,cpb107306}. It is therefore of both fundamental and applied interest to explore the microscopic physical mechanism of the spin relaxation\cite{rmd323,nrm2584,cpb107306}. It is established that the spin relaxation mechanism can be mainly classified into two categories: the D'yakonov-Perel' (DP)\cite{dy110} and Elliot-Yaffet (EY)\cite{pr266,Yafet} mechanisms. For the former, the diffuse scattering changes the direction of an effective magnetic field randomly, which in turn results in the randomisation of spin orientations after a certain diffusion length\cite{spinbook,dy110}. In the case of EY mechanism\cite{pr266,Yafet}, since the eigenstate is a linear combination of spin-up and spin-down states, the diffuse scattering results in a spin-flip scattering responsible for a spin decoherence. Moreover, the EY mechanism plays a crucial role for the spin relaxation in the centrosymmetric system while DP mechanism dominates the spin relaxation in the system without inversion symmetry\cite{pr61,spinbook}.

The DP-type spin relaxation is governed by the momentum $\mathbf{k}$ dependent spin–orbit field $\mathbf{\Omega_k}$, which acts as an effective magnetic field and defines the precession axis and precession frequency\cite{pr61,spinbook}. $\mathbf{\Omega_k}$ is defined from the SOC Hamiltonian $\hbar/2\mathbf{\Omega_k}\cdot{\bm\sigma}$\cite{spinbook,jpd113001,rmp011001}, where $\hbar$ is the reduced Planck's constant and $\bm\sigma=(\sigma_x, \sigma_y, \sigma_z)$ is the vector of Pauli matrices. In addition, the type of $\mathbf{\Omega_k}$ is dictated by the little cogroup of a wave vector in the Brillouin zone\cite{Tinkham,Dresselhaus,rnc609,prbL121125}. The spin relaxation has been widely studied in various nonmagnetic materials based on the spin diffusion theory\cite{prb3912,prb15582,prb155308,prb125307,prb155205,prb174301}. An overall conclusion it that the spin relaxation due to $\mathbf{\Omega_k}$ results in the spin decoherence and a finite spin lifetime. However, the SOC with a persistent spin texture\cite{rmp011001,prl236601,sct073002,nc2763,ma1211,prl216405,apl122903,nc7999,afme75621} is an outstanding exception, where $\mathbf{\Omega_k}$ becomes unidirectional in momentum space and a promising infinite spin lifetime is expected even in the presence of disorder and imperfections\cite{rmp011001,prl236601,njp123005}. 

\begin{figure*}
\includegraphics[width=0.9\textwidth]{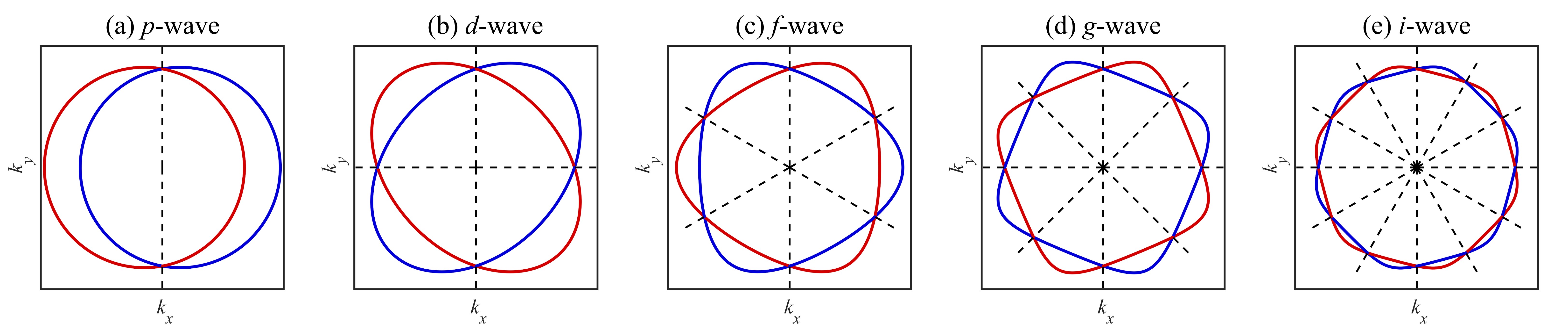}%
\caption{\label{f-1} Schematic Fermi contours for the 2D $p$-wave magnet (a), $d$-wave magnet (b), $f$-wave magnet (c), $g$-wave magnet (d), and $i$-wave magnet (e).  Red and blue contour lines denote two spin channels. The dashed lines indicate node lines along which the spin splitting vanishes.}
\end{figure*}

Recently, a distinct and promising magnetic phase known as the altermagnet was proposed\cite{prx031042,prx040501,afm2409327,jcas2257,nrm2025,nature837}. An altermagnet is characterized by the compensated magnetization and sizable nonrelativistic spin splitting (of the order of $1.0$ eV)\cite{jpsj123702,prb014422,nc2846}. In particular, the nonrelativistic spin splitting is momentum dependent, which is quite similar to the spin splitting due to SOC in nonmagnetic materials\cite{jpd113001,rnc609}. In analogy with $\mathbf{\Omega_k}$, one can define the $\mathbf{k}$ dependent effective magnetic field as $2/\hbar\Delta h(\mathbf{k})\mathbf{\hat{n}}$ from the altermagnetic Hamiltonian term $\Delta h(\mathbf{k})\mathbf{\hat{n}}\cdot\bm{\sigma}$\cite{prx031042,prx040501}, where $\Delta$ indicates the altermagnetic spin-split strength, $h(\mathbf{k})$ is determined by the spin-group symmetry\cite{prx031042,prx040501}, and the unit vector $\mathbf{\hat{n}}=(\hat{n}_x, \hat{n}_y, \hat{n}_z)$ denotes the N\'eel vector direction. As distinct from nonmagnetic materials, $\Delta h(\mathbf{k})\mathbf{\hat{n}}$ is also dependent on $\mathbf{\hat{n}}$, which enables the flexible control of the spin relaxation. Given the similarity to $\mathbf{\Omega_k}$, it is enlightening to examine the spin relaxation in altermagnets. This has been recently explored in the $d$-wave altermagnet without considering SOC using kinetic spin Bloch equations\cite{prb024412}. It is shown that the spin relaxation reveals the DP type in both strong- and weak-scattering regimes\cite{prb024412}. On the other hand, the unconventional $p$-wave and $f$-wave magnets were recently proposed beyond the even-parity altermagnets\cite{01607,00717,na64,na837,prbL220412}, which include the $d$-wave, $g$-wave and $i$-wave altermagnets. For the $p$-wave and $f$-wave magnets, the energy eigenvalues satisfy $\epsilon_{\mathbf{k}s}=\epsilon_{-\mathbf{k}-s}$ ($s=\uparrow, \downarrow$ for spin index, see Fig. \ref{f-1}). On the contrary, it turns out to be $\epsilon_{\mathbf{k}s}=\epsilon_{-\mathbf{k}s}$ for the even-parity altermagnets (see Fig. \ref{f-1}). Those different unconventional magnets can be renamed as $X$-wave magnets with $X=p, d, f, g, i$ irrespective of the parity\cite{prb235307,prb155303,ape030101}. The systematic study on the spin relaxation for all $X$-wave magnets remains unexplored. In particular, the anisotropic spin relaxation rate and the N\'eel vector tunable spin relaxation remains to be studied. It is the purpose of this work to explore the spin relaxation in $X$-wave magnets. 

The rest of the paper is organized as follows. In Sec. \ref{sec2}, we present the theoretical formalism and general formula for spin relaxation calculations. In Sec. \ref{sec3}, we present the results for the spin relaxation in uniform case. Finally, Sec. \ref{sec4} is reserved for further discussion and conclusion.

\section{Theoretical formalism\label{sec2}}

\begin{table*}
\caption{\label{table1} The function $h(\mathbf{k})$ and the tensor of reciprocal spin relaxation times $1/\tau_{ij}$ ($i, j=x, y, z$) for different $X$-wave magnets with the N\'eel vector along the $[0 0 1]$ direction. The constant parameter $C$ is defined as $C=4m^2D/\hbar^4$ while $N_X$ indicates the number of nodes for $h(\mathbf{k})$, as indicated by dashed lines in Fig. \ref{f-1}.}
\begin{ruledtabular}
\begin{tabular}{ccccccccc}
$X$-wave & $h(\mathbf{k})$ & $N_X$ & $1/\tau_{xx}$ & $1/\tau_{yy}$ & $1/\tau_{zz}$ & $1/\tau_{xy}$ & $1/\tau_{xz}$ & $1/\tau_{yz}$\\
\hline
$p$-wave & $k_x$  & $1$ & $C(\Delta^2+\alpha^2)$ & $C(\Delta^2+\alpha^2)$ & $2C\alpha^2$  & $0$ & $0$ & $-C\Delta\alpha$\\
$d$-wave & $k_xk_y$  & $2$ & $C(\Delta^2k_F^2/4+\alpha^2)$ & $C(\Delta^2k_F^2/4+\alpha^2)$ & $2C\alpha^2$  & $0$ & $0$ & $0$\\
$f$-wave & $k_x(k_x^2-3k_y^2)$ & $3$ & $C(\Delta^2k_F^4+\alpha^2)$ & $C(\Delta^2k_F^4+\alpha^2)$ & $2C\alpha^2$  & $0$ & $0$ & $0$\\
$g$-wave & $k_xk_y(k_x^2-k_y^2)$ & $4$ & $C(\Delta^2k_F^6/16+\alpha^2)$ & $C(\Delta^2k_F^6/16+\alpha^2)$ & $2C\alpha^2$  & $0$ & $0$ & $0$\\
$i$-wave & $k_xk_y(3k_x^2-k_y^2)(k_x^2-3k_y^2)$ & $6$ & $C(\Delta^2k_F^{10}/4+\alpha^2)$ & $C(\Delta^2k_F^{10}/4+\alpha^2)$ & $2C\alpha^2$  & $0$ & $0$ & $0$\\
\end{tabular}
\end{ruledtabular}
\end{table*}

We start from the following Hamiltonian describing the two-dimensional (2D) $X$-wave ($X=p, d, f, g, i$) magnets with Rashba SOC 
\begin{equation}\label{eq-1}
    \mathcal{H}=\frac{\hbar^2k^2}{2m}+\alpha(k_x\sigma_y-k_y\sigma_x)+\Delta h(\mathbf{k})\mathbf{\hat{n}}\cdot\bm{\sigma}.
\end{equation}
The first term represents the kinetic energy, where the electron effective mass $m$ is assumed to be isotropic and $\mathbf{k}=(k_x, k_y)$ represents the wave vector. The second term represents the Rashba SOC\cite{r1109,nrp642} and $\alpha$ indicates its strength. The third term represents Zeeman term  induced by the $X$-wave magnet\cite{prx031042,prx040501}, where $\Delta$ indicates spin-split strength and $h(\mathbf{k})$ for different $X$-wave magnets are listed in Table \ref{table1}.

Within the framework of the DP mechanism and relaxation time $\tau$ approximation, the kinetic equation of the averaged spin $\mathbf{S}(\mathbf{r},t)$ is given by\cite{spinbook,rmp011001,prb3912,njp123005}
\begin{widetext}
\begin{equation}\label{eq-2}
    \partial_t\mathbf{S}=-2\tau\langle(\mathbf{v_k}\cdot \nabla)(\mathbf{\Omega_k}\times\mathbf{S})\rangle+\tau\langle(\mathbf{\Omega_k}\cdot\mathbf{S})\mathbf{\Omega_k}\rangle-\tau\langle\Omega_k^2\mathbf{S}\rangle+\tau\langle(\mathbf{v_k}\cdot \nabla)^2\mathbf{S}\rangle,
\end{equation}
where the $\mathbf{k}$-dependent spin-orbit field $\mathbf{\Omega_k}$ resembles an effective magnetic field. From Eq. (\ref{eq-1}), we have $\mathbf{\Omega_k}=2/\hbar(-\alpha k_y+\Delta h_k\hat{n}_x, \alpha k_x+\Delta h_k\hat{n}_y, \Delta h_k\hat{n}_z)$. It is convenient to introduce the Fourier transform: $\mathbf{S}(\mathbf{r},t)=\frac{1}{V}\sum_\mathbf{q}\frac{1}{2\pi}\int d\omega \mathbf{S}(\mathbf{q},\omega)e^{i(\mathbf{q}\cdot \mathbf{r}-\omega t)}$ ($V$ for volume, $\mathbf{q}$ for wave vector and $\omega$ for frequency). We substitute the transform into Eq. (\ref{eq-2}) and find
\begin{equation}\label{eq-3}
    -i\omega\mathbf{S}=-2i\tau\langle(\mathbf{v_k}\cdot\mathbf{q})(\mathbf{\Omega_k}\times\mathbf{S})\rangle+\tau\langle(\mathbf{\Omega_k}\cdot\mathbf{S})\mathbf{\Omega_k}\rangle-\tau\langle\Omega_k^2\mathbf{S}\rangle-\tau\langle(\mathbf{v_k}\cdot\mathbf{q})^2\mathbf{S}\rangle.
\end{equation}
We can rewrite Eq. (\ref{eq-3}) in a matrix form
\begin{equation}\label{eq-4}
   [(-i\omega+Dq^2)I+D_{SO}]\mathbf{S}(\mathbf{q},\omega)=\mathbf{0},
\end{equation}
in which $I$ is the $3\times3$ identity matrix, $D=\hbar^2k_F^2\tau/(2m^2)$ is the diffusion constant ($k_F$ for Fermi wave number), and the matrix $D_{SO}$ reads
\begin{equation}\label{eq-5}
   D_{SO}=\tau\left(
            \begin{array}{ccc}
              \langle\Omega^2-\Omega_x^2\rangle & -\langle\Omega_x\Omega_y\rangle-2i\langle \mathbf{v}\cdot \mathbf{q}\Omega_z\rangle & -\langle\Omega_x\Omega_z\rangle+2i\langle \mathbf{v}\cdot \mathbf{q}\Omega_y\rangle \\
              -\langle\Omega_x\Omega_y\rangle+2i\langle \mathbf{v}\cdot \mathbf{q}\Omega_z\rangle & \langle\Omega^2-\Omega_y^2\rangle & -\langle\Omega_y\Omega_z\rangle-2i\langle \mathbf{v}\cdot \mathbf{q}\Omega_x\rangle \\
              -\langle\Omega_x\Omega_z\rangle-2i\langle \mathbf{v}\cdot \mathbf{q}\Omega_y\rangle & -\langle\Omega_y\Omega_z\rangle+2i\langle \mathbf{v}\cdot \mathbf{q}\Omega_x\rangle & \langle\Omega^2-\Omega_z^2\rangle \\
            \end{array}
          \right).
\end{equation}
\end{widetext}
To simplify the notation, we omit the index $\mathbf{k}$ for $\mathbf{\Omega_k}$ in Eq. (\ref{eq-5}). It is seen that the effects of SOC and $X$-wave magnets are encoded in the matrix $D_{SO}$. For a given $h(\mathbf{k})$ and so $\mathbf{\Omega_k}$, one first calculates $D_{SO}$ from Eq. (\ref{eq-5}). Then, substitute $D_{SO}$ into Eq. (\ref{eq-4}), one can obtain the eigenvalue $\omega$ and eigenstate $\mathbf{S}(\mathbf{q},\omega)$. Finally, the inverse Fourier transform of $\mathbf{S}(\mathbf{q},\omega)$ gives rise to $\mathbf{S}(\mathbf{r},t)$. 

\section{Spin relaxation\label{sec3}}
We can now proceed to calculate $D_{SO}$ for different $X$-wave magnets ($X=p, d, f, g, i$) and the corresponding $h(\mathbf{k})$ are listed in Table \ref{table1}. Details of the derivation of $D_{SO}$ are presented in Appendix \ref{secA}. For the $p$-wave magnet, we have
\begin{widetext}
\begin{equation}\label{eq-6}
    D_{SO}^p=\frac{4m^2D}{\hbar^2}\left(
                   \begin{array}{ccc}
                     \frac{(\alpha+\Delta\hat{n}_y)^2+(\Delta\hat{n}_z)^2}{\hbar^2} & -\frac{\Delta(\alpha\hat{n}_x+\Delta\hat{n}_x\hat{n}_y)}{\hbar^2}-i\frac{\Delta\hat{n}_zq_x}{m} & i\frac{(\alpha+\Delta\hat{n}_y)q_x}{m}-\frac{\Delta^2\hat{n}_x\hat{n}_z}{\hbar^2} \\[1ex]
                     -\frac{\Delta(\alpha\hat{n}_x+\Delta\hat{n}_x\hat{n}_y)}{\hbar^2}+i\frac{\Delta\hat{n}_zq_x}{m} & \frac{\Delta^2(\hat{n}_x^2+\hat{n}_z^2)+\alpha^2}{\hbar^2} & -\frac{\Delta(\alpha+\Delta\hat{n}_y)\hat{n}_z}{\hbar^2}+i\frac{\alpha q_y-\Delta\hat{n}_xq_x}{m}\\[1ex]
                     -i\frac{(\alpha+\Delta\hat{n}_y)q_x}{m}-\frac{\Delta^2\hat{n}_x\hat{n}_z}{\hbar^2} & -\frac{\Delta(\alpha+\Delta\hat{n}_y)\hat{n}_z}{\hbar^2}-i\frac{\alpha q_y-\Delta\hat{n}_xq_x}{m} & \frac{\Delta^2\hat{n}_x^2+(\alpha+\Delta\hat{n}_y)^2+\alpha^2}{\hbar^2}\\[1ex]
                   \end{array}
                 \right),     
\end{equation}
and for the $d$-wave magnet
\begin{equation}\label{eq-7}
 D_{SO}^d=\frac{4m^2D}{\hbar^2}\left(
                   \begin{array}{ccc}
                      \frac{m^2D\Delta^2(\hat{n}_y^2+\hat{n}_z^2)}{2\hbar^4\tau}+\frac{\alpha^2}{\hbar^{2}} & -\frac{m^2D\Delta^2\hat{n}_x \hat{n}_y}{2\hbar^4\tau} & -\frac{m^2D\Delta^2\hat{n}_x \hat{n}_z}{2\hbar^4\tau} + i\frac{\alpha q_x}{m} \\[1ex]
                     -\frac{m^2D\Delta^2\hat{n}_x \hat{n}_y}{2\hbar^4\tau} &  \frac{Dm^2\Delta^2(\hat{n}_x^2+\hat{n}_z^2)}{2\hbar^4\tau}+\frac{\alpha^2}{\hbar^{2}} & -\frac{m^2D\Delta^2\hat{n}_y \hat{n}_z}{2\hbar^4\tau} + i\frac{\alpha q_y}{m} \\[1ex]
                     -\frac{m^2D\Delta^2\hat{n}_x \hat{n}_z}{2\hbar^4\tau} - i\frac{\alpha q_x}{m} & -\frac{m^2D\Delta^2\hat{n}_y \hat{n}_z}{2\hbar^4\tau} - i\frac{\alpha q_y}{m} &  \frac{m^2D\Delta^2(\hat{n}_x^2+\hat{n}_y^2)}{2\hbar^4\tau}+\frac{2\alpha^2}{\hbar^{2}} \\[1ex]
                   \end{array}
                 \right),
\end{equation}
and for the $f$-wave magnet
\begin{equation}\label{eq-8}
   D_{SO}^f=\frac{4m^2D}{\hbar^2}\left(
                   \begin{array}{ccc}
                      \frac{4m^4D^2\Delta^2 (\hat{n}_y^2+\hat{n}_z^2)}{\hbar^6\tau^2}+\frac{\alpha^2}{\hbar^{2}} & -\frac{4m^4D^2\Delta^2\hat{n}_x \hat{n}_y}{\hbar^6\tau^2} & -\frac{4m^4D^2\Delta^2\hat{n}_x \hat{n}_z}{\hbar^6\tau^2} + i\frac{\alpha q_x}{m} \\[1ex]
                     -\frac{4m^4D^2\Delta^2\hat{n}_x \hat{n}_y }{\hbar^6\tau^2} &  \frac{4m^4D^2\Delta^2(\hat{n}_x^2+\hat{n}_z^2)}{\hbar^6\tau^2}+\frac{\alpha^2}{\hbar^{2}} & -\frac{4m^4D^2\Delta^2\hat{n}_y \hat{n}_z}{\hbar^6\tau^2} + i\frac{\alpha q_y}{m} \\[1ex]
                     -\frac{4m^4D^2\Delta^2\hat{n}_x \hat{n}_z}{\hbar^6\tau^2} - i\frac{\alpha q_x}{m} & -\frac{4m^4D^2\Delta^2\hat{n}_y \hat{n}_z}{\hbar^6\tau^2} - i\frac{\alpha q_y}{m} &  \frac{4m^4D^2\Delta^2(\hat{n}_x^2+\hat{n}_y^2)}{\hbar^6\tau^2}+\frac{2\alpha^2}{\hbar^2} \\[1ex]
                   \end{array}
                 \right),
\end{equation}
and for the $g$-wave magnet
\begin{equation}\label{eq-9}
   D_{SO}^g=\frac{4m^2D}{\hbar^2}\left(
                   \begin{array}{ccc}
                      \frac{m^6D^3\Delta^2(\hat{n}_y^2+\hat{n}_z^2)}{2\hbar^8\tau^3}+\frac{\alpha^2}{\hbar^{2}} & -\frac{m^6D^3\Delta^2\hat{n}_x \hat{n}_y}{2\hbar^8\tau^3} & -\frac{m^6D^3\Delta^2\hat{n}_x \hat{n}_z}{2\hbar^8\tau^3} + i\frac{\alpha q_x}{m} \\[1ex]
                     -\frac{m^6D^3\Delta^2\hat{n}_x \hat{n}_y}{2\hbar^8\tau^3} &  \frac{m^6D^3\Delta^2(\hat{n}_x^2+\hat{n}_z^2)}{2\hbar^8\tau^3}+\frac{\alpha^2}{\hbar^{2}} & -\frac{m^6D^3\Delta^2\hat{n}_y \hat{n}_z}{2\hbar^8\tau^3} + i\frac{\alpha q_y}{m} \\[1ex]
                     -\frac{m^6D^3\Delta^2\hat{n}_x \hat{n}_z}{2\hbar^8\tau^3 } - i\frac{\alpha q_x}{m} & -\frac{m^6D^3\Delta^2\hat{n}_y \hat{n}_z}{2\hbar^8\tau^3} - i\frac{\alpha q_y}{m} & \frac{m^6D^3\Delta^2(\hat{n}_x^2+\hat{n}_y^2)}{2\hbar^8\tau^3}+\frac{2\alpha^2}{\hbar^{2}}\\[1ex]
                   \end{array}
                 \right),
\end{equation}
and for the $i$-wave magnet
\begin{equation}\label{eq-10}
D_{SO}^i=\frac{4m^2D}{\hbar^2}\left(
                   \begin{array}{ccc}
                     \frac{8m^{10}D^5\Delta^2(\hat{n}_y^2+\hat{n}_z^2)}{\hbar^{12}\tau^5}+\frac{\alpha^2}{\hbar^{2}} & -\frac{8m^{10}D^5\Delta^2\hat{n}_x \hat{n}_y}{\hbar^{12}\tau^5} & -\frac{8m^{10}D^5\Delta^2\hat{n}_x \hat{n}_z}{\hbar^{12}\tau^5} + i\frac{\alpha q_x}{m} \\[1ex]
                     -\frac{8m^{10}D^5\Delta^2\hat{n}_x \hat{n}_y}{\hbar^{12}\tau^5} & \frac{8m^{10}D^5\Delta^2(\hat{n}_x^2+\hat{n}_z^2)}{\hbar^{12}\tau^5}+\frac{\alpha^2}{\hbar^{2}} & -\frac{8m^{10}D^5\Delta^2\hat{n}_y \hat{n}_z}{\hbar^{12}\tau^5} + i\frac{\alpha q_y}{m} \\[1ex]
                     -\frac{8m^{10}D^5\Delta^2\hat{n}_x \hat{n}_z}{\hbar^{12}\tau^5} - i\frac{\alpha q_x}{m} & -\frac{8m^{10}D^5\Delta^2\hat{n}_y \hat{n}_z}{\hbar^{12}\tau^5} - i\frac{\alpha\hbar^2 q_y}{m} & \frac{8m^{10}D^5\Delta^2(\hat{n}_x^2+\hat{n}_y^2)}{\hbar^{12}\tau^5}+\frac{2\alpha^2}{\hbar^{2}} \\ [1ex]
                   \end{array}
                 \right).
\end{equation}
\end{widetext}

In the following, we consider the special case of $\mathbf{q}=\mathbf{0}$, for which the averaged spin $\mathbf{S}(t)$ reveals uniform distribution and precesses coherently. From Eq. (\ref{eq-4}), we obtain the kinetic equation for $\mathbf{S}(t)$
\begin{equation}\label{eq-11}
   \partial_t\mathbf{S}+D_{SO}\mid_{\mathbf{q}=\mathbf{0}}\mathbf{S}=\mathbf{0},
\end{equation}
Equation \ref{eq-11} describes the relaxation of $\mathbf{S}$ and represents the coupled ordinary first-order differential equations. Thus, $D_{SO}\mid_{\mathbf{q}=0}$ represents the tensor of reciprocal spin relaxation times (spin relaxation rates) $1/\tau_{ij}$ ($i, j=x, y, z$) and reads
\begin{equation}\label{eq-12}
    D_{SO}\mid_{\mathbf{q}=0}=\left(
                     \begin{array}{ccc}
                       1/\tau_{xx} & 1/\tau_{xy} & 1/\tau_{xz} \\
                       1/\tau_{yx} & 1/\tau_{yy} & 1/\tau_{yz} \\
                       1/\tau_{zx} & 1/\tau_{zy} & 1/\tau_{zz} \\
                     \end{array}
                   \right),
\end{equation}
where $\tau_{ij}$ describes the spin relaxation time of $S_i$ component due to the effect of $S_j$ component. As an illustration, we consider the N\'eel vector along the $[0 0 1]$ direction, namely $\mathbf{\hat{n}}=(0, 0, 1)$, from Eqs. (\ref{eq-6})-(\ref{eq-10}), we obtain $1/\tau_{ij}$ for different $X$-wave magnets as follows:
\begin{equation}\label{eq-13}
    \frac{1}{\tau^p}=\frac{4m^2D}{\hbar^4}\left(
                   \begin{array}{ccc}
                     \Delta^2+\alpha^2 & 0 & 0\\
                     0 & \Delta^2+\alpha^2 & -\Delta\alpha\\
                     0 & -\Delta\alpha & 2\alpha^2\\
                   \end{array}
                 \right),    
\end{equation}
and 
\begin{equation}\label{eq-14}
    \frac{1}{\tau^d}=\frac{4m^2D}{\hbar^4}\left(
                   \begin{array}{ccc}
                     \frac{\Delta^2k_F^2}{4}+\alpha^2 & 0 & 0\\
                     0 & \frac{\Delta^2k_F^2}{4}+\alpha^2 & 0\\
                     0 & 0 & 2\alpha^2\\
                   \end{array}
                 \right),    
\end{equation}
and
\begin{equation}\label{eq-15}
    \frac{1}{\tau^f}=\frac{4m^2D}{\hbar^4}\left(
                   \begin{array}{ccc}
                     \Delta^2k_F^4+\alpha^2 & 0 & 0\\
                     0 & \Delta^2k_F^4+\alpha^2 & 0\\
                     0 & 0 & 2\alpha^2\\
                   \end{array}
                 \right),    
\end{equation}
and
\begin{equation}\label{eq-16}
    \frac{1}{\tau^g}=\frac{4m^2D}{\hbar^4}\left(
                   \begin{array}{ccc}
                     \frac{\Delta^2k_F^6}{16}+\alpha^2 & 0 & 0\\
                     0 & \frac{\Delta^2k_F^6}{16}+\alpha^2 & 0\\
                     0 & 0 & 2\alpha^2\\
                   \end{array}
                 \right),    
\end{equation}
and
\begin{equation}\label{eq-17}
    \frac{1}{\tau^i}=\frac{4m^2D}{\hbar^4}\left(
                   \begin{array}{ccc}
                     \frac{\Delta^2k_F^{10}}{4}+\alpha^2 & 0 & 0\\
                     0 & \frac{\Delta^2k_F^{10}}{4}+\alpha^2 & 0\\
                     0 & 0 & 2\alpha^2\\
                   \end{array}
                 \right).    
\end{equation}

It is seen that $1/\tau_{ij}$ for $d, f, g, i$-wave magnets are diagonal, which suggests that the spin relaxations for the three spin components are independent. In addition, $1/\tau_{xx}=1/\tau_{yy}\neq1/\tau_{zz}$ indicates the spin relaxation anisotropy. In the case of the $p$-wave magnet, the off-diagonal components $1/\tau_{yz}$ and $1/\tau_{zy}$ result in the coupling between the $S_y$ and $S_z$ components while the spin relaxation for the $S_x$ component is independent. Those $1/\tau_{ij}$ elements for different $X$-wave magnets are summarized in Table \ref{table1}. From Eq. (\ref{eq-13}), we obtain the spin relaxation equation for the $p$-wave magnet
\begin{equation}\label{eq-18}
\begin{aligned}
&\partial_tS_x=-\frac{S_x}{\tau_{xx}}, \partial_tS_y=-\frac{S_y}{\tau_{yy}}-\frac{S_z}{\tau_{yz}},\\
&\partial_tS_z=-\frac{S_y}{\tau_{zy}}-\frac{S_z}{\tau_{zz}},
\end{aligned}
\end{equation}
where $1/\tau_{ij}$ ($i, j=x, y, z$) read
\begin{equation}\label{eq-19}
\begin{aligned}
&\frac{1}{\tau_{xx}}=\frac{1}{\tau_{yy}}=\frac{2(\Delta^2+\alpha^2)k_F^2\tau}{\hbar^2},\\
&\frac{1}{\tau_{zz}}=\frac{4\alpha^2k_F^2\tau}{\hbar^2}, \frac{1}{\tau_{yz}}=\frac{1}{\tau_{zy}}=-\frac{2\Delta\alpha k_F^2\tau}{\hbar^2}.
\end{aligned}
\end{equation}

We first note that the spin relaxation rates $1/\tau_{xx}$, $1/\tau_{yy}$ and $1/\tau_{zz}$  are proportional to the momentum relaxation time $\tau$, indicative of a strong scattering regime. From Eq. (\ref{eq-18}), the kinetic equations for $S_y$ and $S_z$ represent coupled ordinary first-order differential equations and the general solution can be obtained as
\begin{equation}\label{eq-20}
\begin{aligned}
&S_x=S_x^0\exp[-\frac{2(\Delta^2+\alpha^2)k_F^2\tau}{\hbar^2}t],\\
&S_y=C_1\exp[-\lambda_1t]+C_2\exp[-\lambda_2t],\\
&S_z=C_3\exp[-\lambda_1t]+C_4\exp[-\lambda_2t],\\
\end{aligned}
\end{equation}
where $S_x^0$ is the initial value of $S_x$, that is, $S_x^0=S_x(t=0)$. $\lambda_{1}$ and $\lambda_{2}$ are the eigenvalues of the matrix $1/\tau^p$. For each of $\lambda$'s, there is an eigenvector of $1/\tau^p$,
which we denote by $[C_1\ C_2]^{T}$ and $[C_3\ C_4]^{T}$ ($T$ for transpose). With the use of Eq. (\ref{eq-13}), we have
\begin{equation}\label{eq-21}
\begin{aligned}
&\lambda_1=\frac{2\alpha^2k_F^2\tau}{\hbar^2}, \lambda_2=\frac{2(\Delta^2+2\alpha^2)k_F^2\tau}{\hbar^2},\\
&C_3=\frac{\Delta}{\alpha}C_1, C_4=-\frac{\alpha}{\Delta}C_2,\\
\end{aligned}
\end{equation}
where $C_{1, 2}$ can be determined from the initial values $S_y^0$ and $S_z^0$. Thus,
\begin{equation}\label{eq-22}
C_1=\frac{\alpha^2S_y^0+\Delta\alpha S_z^0}{\Delta^2+\alpha^2}, C_2=\frac{\Delta^2S_y^0-\Delta\alpha S_z^0}{\Delta^2+\alpha^2}.
\end{equation}
\begin{figure}
\includegraphics[width=0.45\textwidth]{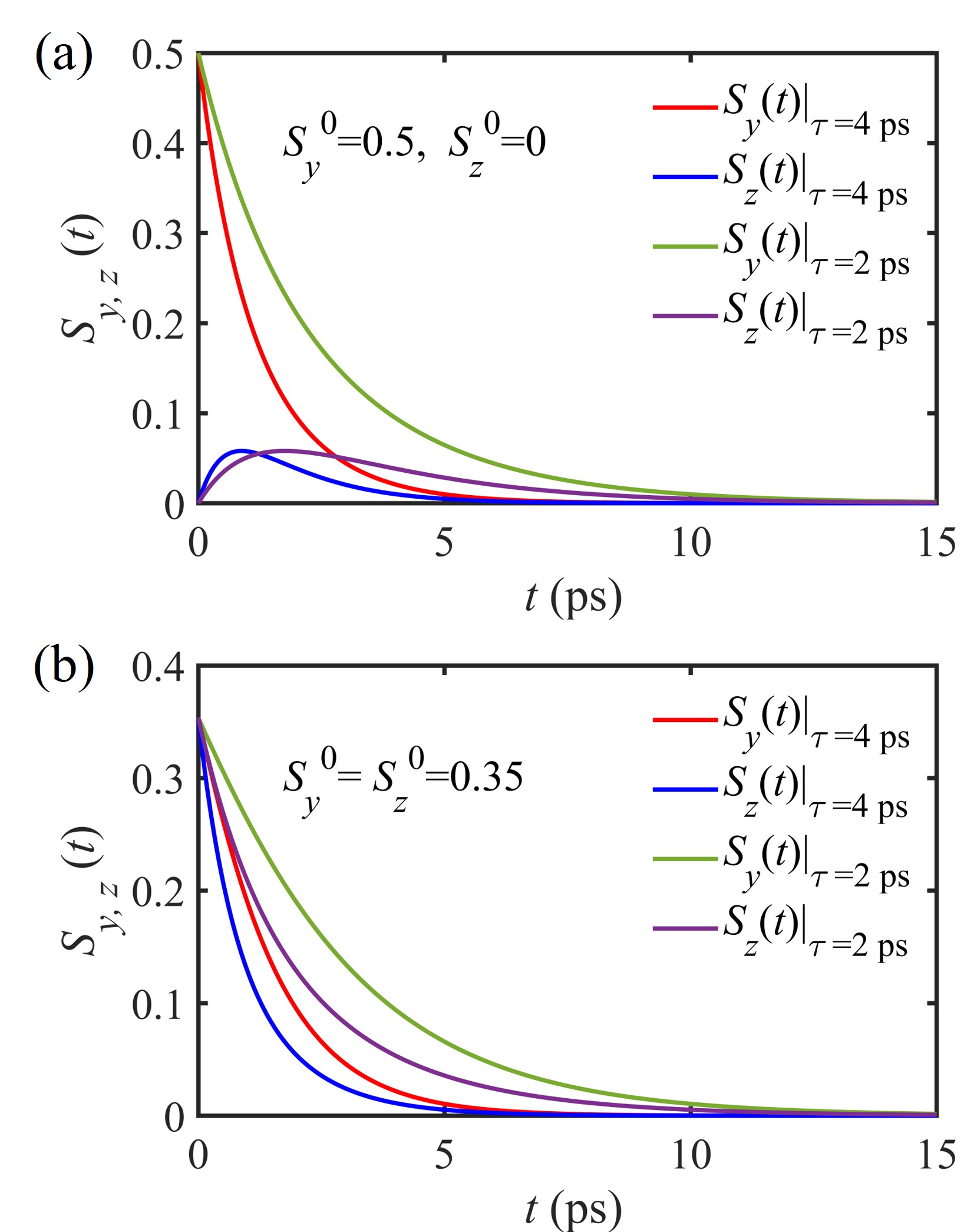}%
\caption{\label{f-2} The temporal evolution of the spin components $S_y$ and $S_z$ for the $p$-wave magnet for $S_y^0=0.5, S_z^0=0$ (a) and $S_y^0=S_z^0=0.35$ (b) with different $\tau$'s. The other parameters are assumed to be $m=0.5$ $m_0$ ($m_0$ for electron rest mass), $k_F=0.1$ Å$^{-1}$, $\alpha=2.0$ meV Å and $\Delta=1.0$ meV Å.}
\end{figure}

Figure \ref{f-2} shows the temporal evolution of the spin components $S_y$ and $S_z$ for the $p$-wave magnet with different initial values $S_y^0$ and $S_z^0$. As seen from Fig. \ref{f-2}(a) for $S_y^0=0.5, S_z^0=0$, $S_y(t)$ decays exponentially in a characteristic time, which is referred to as the spin lifetime. On the contrary, $S_z(t)$ first increases and then decreases exponentially to zero. In addition, the small momentum relaxation time $\tau$ yields the slow decay of $S_y$ or $S_z$ as seen from Eqs. (\ref{eq-20}) and (\ref{eq-21}). For $S_y^0=S_z^0=0.35$ shown Fig. \ref{f-2}(b), a similar exponential decay of $S_y$ or $S_z$ is observed.

For the $d$-wave magnet, from Eq. (\ref{eq-14}), we have 
\begin{equation}\label{eq-23}
   \frac{1}{\tau_{xx}}=\frac{1}{\tau_{yy}}=(\frac{\Delta^2k_F^4}{2\hbar^2}+\frac{2\alpha^2k_F^2}{\hbar^2})\tau, \frac{1}{\tau_{zz}}=\frac{4\alpha^2k_F^2\tau}{\hbar^2}.
\end{equation}
Again, the spin relaxation rates $1/\tau_{xx}$, $1/\tau_{yy}$ and $1/\tau_{zz}$ are proportional to $\tau$. Moreover, as distinct from the $p$-wave magnet, $1/\tau^d$ matrix is diagonal, which indicates that the spin dynamics for $S_x$, $S_y$ and $S_z$ are independent. A similar phenomenon is also observed for $f$-, $g$- and $i$-wave magnets as seen from Eqs. (\ref{eq-15})-(\ref{eq-17}). For comparison, our derived $1/\tau_{xx}$ or $1/\tau_{yy}$ for the $d$-wave magnet is consistent with previous work by noting that $h(\mathbf{k})=k_x^2-k_y^2$ in Ref. \cite{prb024412}, where the $k_xk_y$ axes are rotated by $45^o$ in comparison with our case.

Up to the present, our discussion of the spin relaxation is focused on the N\'eel vector along the $[0 0 1]$ direction. The spin relaxation for an arbitrary $\mathbf{\hat{n}}$ can be in principle studied by solving the coupled ordinary first-order differential equations Eq. (\ref{eq-11}).

\section{Discussion and summary\label{sec4}}
In this work, we focus on the Rashba SOC in $X$-wave magnets. However, the theory can be extended in a straightforward way to be applicable to other SOC types such as Dresselhaus\cite{pr580,prb245159,prb245141}, Weyl\cite{prl136404,prl216402}, persistent\cite{prl236601,sct073002,nc2763,ma1211,prl216405,apl122903,nc7999,afme75621} or more intricate types\cite{jpd113001}. This is done simply by replacing the second term in Eq. (\ref{eq-1}). Second, the spin relaxation is highly controllable by the N\'eel vector as compared to that in nonmagnetic systems, where the spin-orbit field is dictated by the little cogroup symmetry.

In summary, based on the Boltzmann transport theory and within the framework of the DP relaxation mechanism, we have investigated the spin relaxation in the 2D $X$-wave magnets with the Rashba SOC. We analytically derive the matrices of reciprocal spin relaxation times, which suggest the highly anisotropic spin relaxation. As a special case, we examine the spin relaxation for the N\'eel vector along the $[001]$ direction. It is shown that the matrices of reciprocal spin relaxation times are diagonal for the $d$-, $f$-, $g$- and $i$-wave magnets. For the $p$-wave magnet, the off-diagonal components $1/\tau_{yz}$ and $1/\tau_{zy}$ result in the coupling between the $S_y$ and $S_z$ components while the spin relaxation for the $S_x$ component is independent. We further derive the analytical expressions for the temporal evolution of spins. Our work provides a microscopic physical mechanism for the spin relaxation in $X$-wave magnets and paves an avenue for designing the $X$-wave magnet based spintronic devices.

\begin{center}
{\bf ACKNOWLEDGMENTS}
\end{center}

This research was supported by the National Natural Science Foundation of China (Grant No. 12274102).

\begin{center}
{\bf DATA AVAILABILITY}
\end{center}

The data that support the findings of this article are not publicly available. The data are available from the authors upon reasonable request.
\begin{widetext}

\appendix
\section{Derivation of Eqs. (\ref{eq-6})-(\ref{eq-10})\label{secA}}

For the calculation of spin diffusion matrices $D_{p/d/f/g/i}$, we will use the following integral expressions:
\begin{equation}\label{aq-1}
\begin{aligned}
&\langle k_x^2\rangle=\langle k_y^2\rangle=\frac{1}{2\pi}\int_0^{2\pi}k_F^2\sin^2\theta d\theta=\frac{k_F^2}{2}=\frac{m^2D}{\hbar^2\tau},\\
&\langle k_x^4\rangle=\frac{1}{2\pi}\int_{0}^{2\pi}k_F^4\cos^4\theta d\theta=\frac{3k_F^4}{8}=\frac{3m^4D^2}{2\hbar^4\tau^2},\\
&\langle k_x^2k_y^2\rangle=\frac{1}{2\pi}\int_{0}^{2\pi}k_F^{4}\cos^2\theta\sin^2\theta d\theta=\frac{k_F^4}{8}=\frac{m^4D^2}{2\hbar^4\tau^2},\\ 
&\langle k_x^6\rangle=\frac{1}{2\pi}\int_{0}^{2\pi}k_F^6\cos^6\theta d\theta=\frac{5k_F^6}{16}=\frac{5m^6D^3}{2\hbar^6\tau^3},\\
&\langle k_x^2k_y^4\rangle=\langle k_x^4k_y^2\rangle=\frac{1}{2\pi}\int_{0}^{2\pi}k_F^{6}\cos^4\theta\sin^2\theta d\theta=\frac{k_F^6}{16}=\frac{m^6D^3}{2\hbar^6\tau^3},\\
&\langle k_x^2k_y^6\rangle=\langle k_x^6k_y^2\rangle=\frac{1}{2\pi}\int_{0}^{2\pi}k_F^{8}\cos^6\theta\sin^2\theta d\theta=\frac{5k_F^8}{128}=\frac{5m^8D^4}{8\hbar^8\tau^4},\\ 
&\langle k_x^4k_y^4\rangle=\frac{1}{2\pi}\int_{0}^{2\pi}k_F^{8}\cos^4\theta\sin^4\theta d\theta=\frac{3k_F^8}{128}=\frac{3m^8D^4}{8\hbar^8\tau^4},\\
&\langle k_x^2k_y^{10}\rangle=\langle k_x^{10}k_y^2\rangle=\frac{1}{2\pi}\int_{0}^{2\pi}k_F^{12}\cos^{10}\theta\sin^2\theta d\theta=\frac{21k_F^{12}}{1024}=\frac{21m^{12}D^6}{16\hbar^{12}\tau^6},\\
&\langle k_x^4k_y^8\rangle=\langle k_x^8k_y^4\rangle=\frac{1}{2\pi}\int_{0}^{2\pi}k_F^{12}\cos^8\theta\sin^4\theta d\theta=\frac{7k_F^{12}}{1024}=\frac{7m^{12}D^6}{16\hbar^{12}\tau^6},\\
&\langle k_x^6k_y^6\rangle=\frac{1}{2\pi}\int_{0}^{2\pi}k_F^{12}\cos^6\theta\sin^6\theta d\theta=\frac{5k_F^{12}}{1024}=\frac{5m^{12}D^6}{16\hbar^{12}\tau^6},\\
&\langle k_xk_y\rangle=\langle k_xk_y^2\rangle=\langle k_x^2k_y\rangle=\langle k_xk_y^3\rangle=\langle k_x^3k_y\rangle=\langle k_xk_y^4\rangle=\langle k_x^4k_y\rangle=\langle k_x^2k_y^3\rangle=\langle k_x^3k_y^2\rangle=0,\\
&\langle k_xk_y^6\rangle=\langle k_x^6k_y\rangle=\langle k_x^3k_y^4\rangle=\langle k_x^4k_y^3\rangle=\langle k_x^2k_y^5\rangle=\langle k_x^5k_y^2\rangle=0.\\
\end{aligned}
\end{equation}
For the $p$-wave magnet, we have $\mathbf{\Omega}_{\mathbf{k}}=2/\hbar(\Delta k_x\hat{n}_x-\alpha k_y, \Delta k_x\hat{n}_y+\alpha k_x, \Delta k_x\hat{n}_z)$ and
\begin{equation}\label{aq-2}
\begin{aligned}
&\tau\langle\Omega_x^2\rangle=\frac{4\tau}{\hbar^2}\bigl\langle (\Delta k_x\hat{n}_x-\alpha k_y)^2\bigr\rangle=\frac{4\tau}{\hbar^2}\bigl(\Delta^2\hat{n}_x^2\langle k_x^2\rangle+\alpha^2\langle k_y^2\rangle-2\Delta\hat{n}_x\alpha\langle k_x k_y\rangle\bigr)=\frac{4 m^2D(\Delta^2\hat{n}_x^2 + \alpha^2)}{\hbar^4},\\
&\tau\langle\Omega_y^2\rangle=\frac{4\tau}{\hbar^2}\bigl\langle(\Delta k_x\hat{n}_y+\alpha k_x)^2 \bigr\rangle=\frac{4\tau}{\hbar^2}(\Delta \hat{n}_y + \alpha)^2\langle k_x^2\rangle=\frac{4 m^2 D (\Delta \hat{n}_y+\alpha)^2}{\hbar^4},\\
&\tau\langle\Omega_z^2\rangle=\frac{4\tau}{\hbar^2}\bigl\langle(\Delta k_x\hat{n}_z)^2\bigr\rangle=\frac{4\tau}{\hbar^2}\Delta^2\hat{n}_z^2\langle k_x^2\rangle =\frac{4 m^2 D \Delta^2 \hat{n}_z^2}{\hbar^4},\\
&\tau\langle\Omega_x\Omega_y\rangle = \frac{4\tau}{\hbar^2}\bigl(\Delta^2 \hat{n}_x \hat{n}_y\langle k_x^2\rangle + \Delta \hat{n}_x\alpha\langle k_x^2\rangle - \Delta \hat{n}_y\alpha\langle k_x k_y\rangle - \alpha^2\langle k_x k_y\rangle\bigr)=\frac{4 m^2 D \Delta \hat{n}_x (\Delta \hat{n}_y + \alpha)}{\hbar^4},\\
&\tau\langle\Omega_x\Omega_z\rangle = \frac{4\tau}{\hbar^2}\bigl\langle (\Delta k_x \hat{n}_x - \alpha k_y)\Delta k_x \hat{n}_z \bigr\rangle = \frac{4\tau}{\hbar^2}\Delta \hat{n}_z\bigl(\Delta \hat{n}_x\langle k_x^2\rangle - \alpha\langle k_x k_y\rangle\bigr)=\frac{4 m^2 D \Delta^2 \hat{n}_x \hat{n}_z}{\hbar^4},\\
&\tau\langle\Omega_y\Omega_z\rangle = \frac{4\tau}{\hbar^2}\bigl\langle (\Delta k_x \hat{n}_y + \alpha k_x)\Delta k_x \hat{n}_z \bigr\rangle = \frac{4\tau}{\hbar^2}\Delta \hat{n}_z(\Delta \hat{n}_y + \alpha)\langle k_x^2\rangle = \frac{4 m^2 D \Delta \hat{n}_z (\Delta \hat{n}_y + \alpha)}{\hbar^4},\\
&\tau\langle(\mathbf{v}\cdot\mathbf{q})\Omega_x\rangle = \frac{2\tau}{m}\bigl\langle (k_x q_x + k_y q_y)(\Delta k_x \hat{n}_x - \alpha k_y) \bigr\rangle = \frac{2\tau}{m}\bigl(\Delta \hat{n}_x q_x\langle k_x^2\rangle - \alpha q_y\langle k_y^2\rangle\bigr) = \frac{2 m D(\Delta \hat{n}_x q_x - \alpha q_y)}{\hbar^2},\\ 
&\tau\langle(\mathbf{v}\cdot\mathbf{q})\Omega_y\rangle = \frac{2\tau}{m}\bigl\langle (k_x q_x + k_y q_y)(\Delta k_x \hat{n}_y + \alpha k_x) \bigr\rangle = \frac{2\tau}{m}(\Delta \hat{n}_y + \alpha)q_x\langle k_x^2\rangle= \frac{2 m D (\Delta \hat{n}_y + \alpha)q_x}{\hbar^2},\\
&\tau\langle(\mathbf{v}\cdot\mathbf{q})\Omega_z\rangle = \frac{2\tau}{m}\bigl\langle (k_x q_x + k_y q_y)\Delta k_x \hat{n}_z \bigr\rangle = \frac{2\tau}{m}\Delta \hat{n}_z q_x\langle k_x^2\rangle = \frac{2 m D \Delta \hat{n}_z q_x}{\hbar^2}.\\
\end{aligned}
\end{equation}
For the $d$-wave magnet, we have $\mathbf{\Omega}_{\mathbf{k}} = 2/\hbar(\Delta k_x k_y \hat{n}_x - \alpha k_y, \Delta k_x k_y \hat{n}_y + \alpha k_x, \Delta k_x k_y \hat{n}_z)$ and
\begin{equation}\label{aq-3}
\begin{aligned}
&\tau\langle\Omega_x^2\rangle = \frac{4\tau}{\hbar^2}\bigl\langle (\Delta k_x k_y \hat{n}_x - \alpha k_y)^2 \bigr\rangle = \frac{4\tau}{\hbar^2}\Bigl( \Delta^2 \hat{n}_x^2 \langle k_x^2 k_y^2\rangle - 2 \Delta \hat{n}_x \alpha \langle k_x k_y^2\rangle + \alpha^2 \langle k_y^2\rangle \Bigr) = \frac{4 m^2 D\alpha^2}{\hbar^4} + \frac{2 m^4 D^2 \Delta^2 \hat{n}_x^2}{\hbar^6 \tau},\\
&\tau\langle\Omega_y^2\rangle = \frac{4\tau}{\hbar^2}\bigl\langle (\Delta k_x k_y \hat{n}_y + \alpha k_x)^2 \bigr\rangle  = \frac{4\tau}{\hbar^2}\Bigl( \Delta^2 \hat{n}_y^2 \langle k_x^2 k_y^2\rangle + 2 \Delta \hat{n}_y \alpha \langle k_x^2 k_y\rangle + \alpha^2 \langle k_x^2\rangle \Bigr) = \frac{4 m^2 D\alpha^2}{\hbar^4} + \frac{2 m^4 D^2 \Delta^2\hat{n}_y^2}{\hbar^6 \tau},\\
&\tau\langle\Omega_z^2\rangle = \frac{4\tau}{\hbar^2}\bigl\langle (\Delta k_x k_y \hat{n}_z)^2 \bigr\rangle = \frac{4\tau}{\hbar^2} \Delta^2 \hat{n}_z^2 \langle k_x^2 k_y^2\rangle = \frac{2 m^4 D^2 \Delta^2 \hat{n}_z^2}{\hbar^6 \tau},\\
&\tau\langle\Omega_x\Omega_y\rangle = \frac{4\tau}{\hbar^2}\Bigl( \Delta^2 \hat{n}_x \hat{n}_y \langle k_x^2 k_y^2\rangle + \Delta \hat{n}_x \alpha \langle k_x^2 k_y\rangle - \Delta \hat{n}_y \alpha \langle k_x k_y^2\rangle - \alpha^2 \langle k_x k_y\rangle \Bigr) = \frac{2 m^4 D^2 \Delta^2\hat{n}_x\hat{n}_y}{\hbar^6 \tau},\\
&\tau\langle\Omega_x\Omega_z\rangle = \frac{4\tau}{\hbar^2}\bigl\langle (\Delta k_x k_y \hat{n}_x - \alpha k_y)\,\Delta k_x k_y \hat{n}_z \bigr\rangle  = \frac{4\tau}{\hbar^2}\Bigl( \Delta^2 \hat{n}_x \hat{n}_z \langle k_x^2 k_y^2\rangle - \Delta \hat{n}_z \alpha \langle k_x k_y^2\rangle \Bigr) = \frac{2m^4 D^2 \Delta^2 \hat{n}_x \hat{n}_z}{\hbar^6 \tau},\\
&\tau\langle\Omega_y\Omega_z\rangle = \frac{4\tau}{\hbar^2}\bigl\langle (\Delta k_x k_y \hat{n}_y + \alpha k_x)\,\Delta k_x k_y \hat{n}_z \bigr\rangle = \frac{4\tau}{\hbar^2}\Bigl( \Delta^2 \hat{n}_y \hat{n}_z \langle k_x^2 k_y^2\rangle + \Delta \hat{n}_z \alpha \langle k_x^2 k_y\rangle \Bigr) = \frac{2 m^4 D^2 \Delta^2\hat{n}_y \hat{n}_z}{\hbar^6 \tau},\\ 
&\tau\langle(\mathbf{v}\cdot\mathbf{q})\Omega_x\rangle = \frac{2\tau}{m}\Bigl( \Delta \hat{n}_x q_x \langle k_x^2 k_y\rangle + \Delta \hat{n}_x q_y \langle k_x k_y^2\rangle - \alpha q_x \langle k_x k_y\rangle - \alpha q_y \langle k_y^2\rangle \Bigr) = -\frac{2 m D \alpha q_y}{\hbar^2},\\
&\tau\langle(\mathbf{v}\cdot\mathbf{q})\Omega_y\rangle = \frac{2\tau}{m}\Bigl( \Delta \hat{n}_y q_x \langle k_x^2 k_y\rangle + \Delta \hat{n}_y q_y \langle k_x k_y^2\rangle + \alpha q_x \langle k_x^2\rangle + \alpha q_y \langle k_x k_y\rangle \Bigr) = \frac{2 m D\alpha q_x}{\hbar^2}, \\
&\tau\langle(\mathbf{v}\cdot\mathbf{q})\Omega_z\rangle  = \frac{2\tau}{m} \Delta \hat{n}_z \Bigl( q_x \langle k_x^2 k_y\rangle + q_y \langle k_x k_y^2\rangle \Bigr) = 0.\\
\end{aligned}
\end{equation}
For the $f$-wave magnet, we have $\mathbf{\Omega}_{\mathbf{k}} = 2/\hbar(\Delta k_x (k_x^2 - 3k_y^2) \hat{n}_x - \alpha k_y, \Delta k_x (k_x^2 - 3k_y^2) \hat{n}_y + \alpha k_x, \Delta k_x (k_x^2 - 3k_y^2) \hat{n}_z)$ and
\begin{equation}\label{aq-4}
\begin{aligned}
&\tau\langle\Omega_x^2\rangle = \frac{4\tau}{\hbar^2}\Bigl[ \Delta^2 \hat{n}_x^2 \bigl(\langle k_x^6\rangle - 6\langle k_x^4 k_y^2\rangle + 9\langle k_x^2 k_y^4\rangle\bigr) - 2\Delta \hat{n}_x\alpha \bigl(\langle k_x^3 k_y\rangle - 3\langle k_x k_y^3\rangle\bigr) + \alpha^2 \langle k_y^2\rangle \Bigr] = \frac{4 m^2 D \alpha^2 }{\hbar^4}\,+ \frac{16 m^6 D^3 \Delta^2\hat{n}_x^2}{\hbar^8\tau^2},\\
&\tau\langle\Omega_y^2\rangle = \frac{4\tau}{\hbar^2}\Bigl[ \Delta^2 \hat{n}_y^2 \bigl(\langle k_x^6\rangle - 6\langle k_x^4 k_y^2\rangle + 9\langle k_x^2 k_y^4\rangle\bigr) + 2\Delta \hat{n}_y\alpha \bigl(\langle k_x^4\rangle - 3\langle k_x^2 k_y^2\rangle\bigr) + \alpha^2 \langle k_x^2\rangle \Bigr] = \frac{4 m^2 D\alpha^2}{\hbar^4} + \frac{16 m^6 D^3 \Delta^2 \hat{n}_y^2}{\hbar^8\tau^2},\\
&\tau\langle\Omega_z^2\rangle = \frac{4\tau}{\hbar^2}\bigl\langle \bigl( \Delta k_x(k_x^2-3k_y^2)\hat{n}_z \bigr)^2 \bigr\rangle = \frac{4\tau}{\hbar^2} \Delta^2 \hat{n}_z^2 \bigl(\langle k_x^6\rangle - 6\langle k_x^4 k_y^2\rangle + 9\langle k_x^2 k_y^4\rangle\bigr) 
    = \frac{16 m^6 D^3 \Delta^2 \hat{n}_z^2}{\hbar^8\tau^2},\\
&\tau\langle\Omega_x\Omega_y\rangle = \frac{4\tau}{\hbar^2}\bigl\langle \bigl( \Delta k_x(k_x^2-3k_y^2)\hat{n}_x - \alpha k_y \bigr)\bigl( \Delta k_x(k_x^2-3k_y^2)\hat{n}_y + \alpha k_x \bigr) \bigr\rangle = \frac{16 m^6 D^3 \Delta^2\hat{n}_x\hat{n}_y}{\hbar^8\tau^2},\\
&\tau\langle\Omega_x\Omega_z\rangle = \frac{4\tau}{\hbar^2}\Bigl[ \Delta^2 \hat{n}_x \hat{n}_z \bigl(\langle k_x^6\rangle - 6\langle k_x^4 k_y^2\rangle + 9\langle k_x^2 k_y^4\rangle\bigr) - \Delta \hat{n}_z\alpha \bigl(\langle k_x^3 k_y\rangle - 3\langle k_x k_y^3\rangle\bigr) \Bigr] = \frac{16 m^6 D^3 \Delta^2\hat{n}_x \hat{n}_z}{\hbar^8\tau^2},\\
&\tau\langle\Omega_y\Omega_z\rangle = \frac{4\tau}{\hbar^2}\Bigl[ \Delta^2 \hat{n}_y \hat{n}_z \bigl(\langle k_x^6\rangle - 6\langle k_x^4 k_y^2\rangle + 9\langle k_x^2 k_y^4\rangle\bigr) + \Delta \hat{n}_z\alpha \bigl(\langle k_x^4\rangle - 3\langle k_x^2k_y^2\rangle\bigr) \Bigr] = \frac{16 m^6 D^3 \Delta^2\hat{n}_y \hat{n}_z}{\hbar^8\tau^2},\\
&\tau\langle(\mathbf{v}\cdot\mathbf{q})\Omega_x\rangle = \frac{2\tau}{m}\Bigl[ \Delta \hat{n}_x q_x \bigl(\langle k_x^4\rangle - 3\langle k_x^2 k_y^2\rangle\bigr) + \Delta \hat{n}_x q_y \bigl(\langle k_x^3 k_y\rangle - 3\langle k_x k_y^3\rangle\bigr) - \alpha q_x \langle k_x k_y\rangle - \alpha q_y \langle k_y^2\rangle \Bigr] = -\frac{2 m D \alpha q_y}{\hbar^2},\\
&\tau\langle(\mathbf{v}\cdot\mathbf{q})\Omega_y\rangle = \frac{2\tau}{m}\Bigl[ \Delta \hat{n}_y q_x \bigl(\langle k_x^4\rangle - 3\langle k_x^2 k_y^2\rangle\bigr) + \Delta \hat{n}_y q_y \bigl(\langle k_x^3 k_y\rangle - 3\langle k_x k_y^3\rangle\bigr) + \alpha q_x \langle k_x^2\rangle + \alpha q_y \langle k_x k_y\rangle \Bigr] = \frac{2 m D \alpha q_x}{\hbar^2},\\
&\tau\langle(\mathbf{v}\cdot\mathbf{q})\Omega_z\rangle = \frac{2\tau}{m} \Delta \hat{n}_z \Bigl[ q_x \bigl(\langle k_x^4\rangle - 3\langle k_x^2 k_y^2\rangle\bigr) + q_y \bigl(\langle k_x^3 k_y\rangle - 3\langle k_x k_y^3\rangle\bigr) \Bigr] = 0.\\
\end{aligned}
\end{equation}
For the $g$-wave magnet, we have $\mathbf{\Omega}_{\mathbf{k}} = 2/\hbar( \Delta k_x k_y (k_x^2 - k_y^2) \hat{n}_x - \alpha k_y, \Delta k_x k_y (k_x^2 - k_y^2) \hat{n}_y + \alpha k_x, \Delta k_x k_y (k_x^2 - k_y^2) \hat{n}_z)$ and
\begin{equation}\label{aq-5}
\begin{aligned}
&\tau\langle\Omega_x^2\rangle = \frac{4\tau}{\hbar^2}\Bigl[ \Delta^2 \hat{n}_x^2 \bigl( \langle k_x^6 k_y^2\rangle - 2\langle k_x^4 k_y^4\rangle + \langle k_x^2 k_y^6\rangle \bigr) - 2\Delta \hat{n}_x\alpha \bigl( \langle k_x^3 k_y^2\rangle - \langle k_x k_y^4\rangle \bigr) + \alpha^2 \langle k_y^2\rangle \Bigr] = \frac{4 m^2 D \alpha^2}{\hbar^4}\, + \frac{2 m^8 D^4 \Delta^2\hat{n}_x^2}{\hbar^{10}\tau^3},\\
&\tau\langle\Omega_y^2\rangle = \frac{4\tau}{\hbar^2}\Bigl[ \Delta^2 \hat{n}_y^2 \bigl( \langle k_x^6 k_y^2\rangle - 2\langle k_x^4 k_y^4\rangle + \langle k_x^2 k_y^6\rangle \bigr) + 2\Delta \hat{n}_y\alpha \bigl( \langle k_x^4 k_y\rangle - \langle k_x^2 k_y^3\rangle \bigr) + \alpha^2 \langle k_x^2\rangle \Bigr] = \frac{4 m^2 D \alpha^2}{\hbar^4}\, + \frac{2 m^8 D^4 \Delta^2\hat{n}_y^2}{\hbar^{10}\tau^3},\\
&\tau\langle\Omega_z^2\rangle = \frac{4\tau}{\hbar^2} \Delta^2 \hat{n}_z^2 \bigl( \langle k_x^6 k_y^2\rangle - 2\langle k_x^4 k_y^4\rangle + \langle k_x^2 k_y^6\rangle \bigr) = \frac{2 m^8 D^4 \Delta^2\hat{n}_z^2}{\hbar^{10}\tau^3},\\
&\tau\langle\Omega_x\Omega_y\rangle = \frac{4\tau}{\hbar^2}\bigl\langle \bigl( \Delta k_x k_y(k_x^2-k_y^2)\hat{n}_x - \alpha k_y \bigr)\bigl( \Delta k_x k_y(k_x^2-k_y^2)\hat{n}_y + \alpha k_x \bigr) \bigr\rangle =\frac{2m^8D^4\Delta^2\hat{n}_x\hat{n}_y}{\hbar^{10}\tau^3},\\
&\tau\langle\Omega_x\Omega_z\rangle = \frac{4\tau}{\hbar^2}\Bigl[ \Delta^2 \hat{n}_x \hat{n}_z \bigl( \langle k_x^6 k_y^2\rangle - 2\langle k_x^4 k_y^4\rangle + \langle k_x^2 k_y^6\rangle \bigr) - \Delta \hat{n}_z\alpha \bigl( \langle k_x^3 k_y^2\rangle - \langle k_x k_y^4\rangle \bigr) \Bigr] = \frac{2 m^8 D^4 \Delta^2\hat{n}_x \hat{n}_z}{\hbar^{10}\tau^3},\\
&\tau\langle\Omega_y\Omega_z\rangle = \frac{4\tau}{\hbar^2}\Bigl[ \Delta^2 \hat{n}_y \hat{n}_z \bigl( \langle k_x^6 k_y^2\rangle - 2\langle k_x^4 k_y^4\rangle + \langle k_x^2 k_y^6\rangle \bigr) + \Delta \hat{n}_z\alpha \bigl( \langle k_x^4 k_y\rangle - \langle k_x^2 k_y^3\rangle \bigr) \Bigr] = \frac{2 m^8 D^4 \Delta^2\hat{n}_y \hat{n}_z}{\hbar^{10}\tau^3}.\\
&\tau\langle(\mathbf{v}\cdot\mathbf{q})\Omega_x\rangle=\frac{2\tau}{m}\Bigl[ \Delta \hat{n}_x q_x \bigl( \langle k_x^4 k_y\rangle - \langle k_x^2 k_y^3\rangle \bigr) + \Delta \hat{n}_x q_y \bigl( \langle k_x^3 k_y^2\rangle - \langle k_x k_y^4\rangle \bigr) - \alpha q_x \langle k_x k_y\rangle - \alpha q_y \langle k_y^2\rangle \Bigr] = -\frac{2 m D \alpha q_y}{\hbar^2},\\
&\tau\langle(\mathbf{v}\cdot\mathbf{q})\Omega_y\rangle = \frac{2\tau}{m}\Bigl[ \Delta \hat{n}_y q_x \bigl( \langle k_x^4 k_y\rangle - \langle k_x^2 k_y^3\rangle \bigr) + \Delta \hat{n}_y q_y \bigl( \langle k_x^3 k_y^2\rangle - \langle k_x k_y^4\rangle \bigr) + \alpha q_x \langle k_x^2\rangle + \alpha q_y \langle k_x k_y\rangle \Bigr] = \frac{2 m D \alpha q_x}{\hbar^2},\\
&\tau\langle(\mathbf{v}\cdot\mathbf{q})\Omega_z\rangle = \frac{2\tau}{m} \Delta \hat{n}_z \Bigl[ q_x \bigl( \langle k_x^4 k_y\rangle - \langle k_x^2 k_y^3\rangle \bigr) + q_y \bigl( \langle k_x^3 k_y^2\rangle - \langle k_x k_y^4\rangle \bigr) \Bigr] = 0.
\end{aligned}
\end{equation}
For the $i$-wave magnet, we have $\mathbf{\Omega}_{\mathbf{k}} = 2/\hbar(\Delta k_x k_y (3k_x^2 - k_y^2)(k_x^2 - 3k_y^2) \hat{n}_x - \alpha k_y, \Delta k_x k_y (3k_x^2 - k_y^2)(k_x^2 - 3k_y^2) \hat{n}_y + \alpha k_x, \Delta k_x k_y (3k_x^2 - k_y^2)(k_x^2 - 3k_y^2) \hat{n}_z)$ and
\begin{equation}\label{aq-6}
\begin{aligned}
&\tau\langle\Omega_x^2\rangle = \frac{4\tau}{\hbar^2}\bigl\langle \bigl[ \Delta k_x k_y(3k_x^2-k_y^2)(k_x^2-3k_y^2)\hat{n}_x - \alpha k_y \bigr]^2 \bigr\rangle = \frac{4 m^2 D \alpha^2}{\hbar^4} + \frac{32 m^{12} D^6 \Delta^2\hat{n}_x^2}{\hbar^{14}\tau^5},\\
&\tau\langle\Omega_y^2\rangle = \frac{4\tau}{\hbar^2}\bigl\langle \bigl[ \Delta k_x k_y(3k_x^2-k_y^2)(k_x^2-3k_y^2)\hat{n}_y + \alpha k_x \bigr]^2 \bigr\rangle = \frac{4 m^2 D \alpha^2}{\hbar^4} + \frac{32 m^{12} D^6 \Delta^2\hat{n}_y^2}{\hbar^{14}\tau^5},\\
&\tau\langle\Omega_z^2\rangle = \frac{4\tau}{\hbar^2}\bigl\langle \bigl[ \Delta k_x k_y(3k_x^2-k_y^2)(k_x^2-3k_y^2)\hat{n}_z \bigr]^2 \bigr\rangle = \frac{32 m^{12} D^6 \Delta^2\hat{n}_z^2}{\hbar^{14}\tau^5},\\
&\tau\langle\Omega_x\Omega_y\rangle = \frac{4\tau}{\hbar^2}\bigl\langle \bigl[ \Delta k_x k_y(3k_x^2-k_y^2)(k_x^2-3k_y^2)\hat{n}_x - \alpha k_y \bigr]\bigl[ \Delta k_x k_y(3k_x^2-k_y^2)(k_x^2-3k_y^2)\hat{n}_y + \alpha k_x \bigr] \bigr\rangle = \frac{32 m^{12} D^6 \Delta^2\hat{n}_x\hat{n}_y}{\hbar^{14}\tau^5},\\
&\tau\langle\Omega_x\Omega_z\rangle = \frac{4\tau}{\hbar^2}\bigl\langle\bigl[\Delta k_x k_y(3k_x^2-k_y^2)(k_x^2-3k_y^2)\hat{n}_x - \alpha k_y \bigr][\Delta k_x k_y(3k_x^2-k_y^2)(k_x^2-3k_y^2)\hat{n}_z]\bigr\rangle=\frac{32 m^{12} D^6 \Delta^2\hat{n}_x \hat{n}_z}{\hbar^{14}\tau^5},\\
&\tau\langle\Omega_y\Omega_z\rangle = \frac{4\tau}{\hbar^2}\bigl\langle \bigl[ \Delta k_x k_y(3k_x^2-k_y^2)(k_x^2-3k_y^2)\hat{n}_y + \alpha k_x \bigr]\Delta k_x k_y(3k_x^2-k_y^2)(k_x^2-3k_y^2)\hat{n}_z\bigr\rangle = \frac{32 m^{12} D^6 \Delta^2\hat{n}_y\hat{n}_z}{\hbar^{14}\tau^5},\\
&\tau\langle(\mathbf{v}\cdot\mathbf{q})\Omega_x\rangle = \frac{2\tau}{m}\bigl\langle (k_x q_x + k_y q_y)\bigl[ \Delta k_x k_y(3k_x^2-k_y^2)(k_x^2-3k_y^2)\hat{n}_x - \alpha k_y \bigr] \bigr\rangle= -\frac{2 m D \alpha q_y}{\hbar^2},\\
&\tau\langle(\mathbf{v}\cdot\mathbf{q})\Omega_y\rangle = \frac{2\tau}{m}\bigl\langle (k_x q_x + k_y q_y)\bigl[ \Delta k_x k_y(3k_x^2-k_y^2)(k_x^2-3k_y^2)\hat{n}_y + \alpha k_x \bigr] \bigr\rangle=\frac{2 m D \alpha q_x}{\hbar^2},\\
&\tau\langle(\mathbf{v}\cdot\mathbf{q})\Omega_z\rangle = \frac{2\tau}{m}\bigl\langle (k_x q_x + k_y q_y) \Delta k_x k_y(3k_x^2-k_y^2)(k_x^2-3k_y^2)\hat{n}_z \bigr\rangle=0.\\
\end{aligned}
\end{equation}
\end{widetext}

\end{document}